# In-situ MOCVD Growth and Band Offsets of $Al_2O_3$ Dielectric on β-$Ga_2O_3$ and β-$(Al_xGa_{1-x})_2O_3$ thin films

A F M Anhar Uddin Bhuiyan[1,a)], Lingyu Meng[1], Hsien-Lien Huang[2], Jinwoo Hwang[2], and Hongping Zhao[1,2,b)]

[1]*Department of Electrical and Computer Engineering, The Ohio State University, Columbus, OH 43210, USA*

[2]*Department of Materials Science and Engineering, The Ohio State University, Columbus, OH 43210, USA*

[a)]Email: bhuiyan.13@osu.edu [b)]Corresponding author Email: zhao.2592@osu.edu

## Abstract

The in-situ metalorganic chemical vapor deposition (MOCVD) growth of $Al_2O_3$ dielectrics on β-$Ga_2O_3$ and β-$(Al_xGa_{1-x})_2O_3$ films is investigated as a function of crystal orientations and Al compositions of β-$(Al_xGa_{1-x})_2O_3$ films. The interface and film qualities of $Al_2O_3$ dielectrics are evaluated by high resolution X-ray diffraction (HR-XRD) and scanning transmission electron microscopy (HR-STEM) imaging, which indicate the growth of high quality amorphous $Al_2O_3$ dielectrics with abrupt interfaces on (010), (100) and ($\bar{2}$01) oriented β-$(Al_xGa_{1-x})_2O_3$ films. The surface stoichiometries of $Al_2O_3$ deposited on all orientations of β-$(Al_xGa_{1-x})_2O_3$ are found to be well maintained with a bandgap energy of 6.91 eV as evaluated by high resolution x-ray photoelectron spectroscopy, which is consistent with the atomic layer deposited (ALD) $Al_2O_3$ dielectrics. The evolution of band offsets at both in-situ MOCVD and ex-situ ALD deposited $Al_2O_3$/β-$(Al_xGa_{1-x})_2O_3$ are determined as a function of Al composition, indicating the influence of the deposition method, orientation, and Al composition of β-$(Al_xGa_{1-x})_2O_3$ films on resulting band alignments. Type II band alignments are determined at the MOCVD grown $Al_2O_3$/β-$(Al_xGa_{1-x})_2O_3$ interfaces for (010) and (100) orientations, whereas type I band alignments with relatively lower conduction band offsets are observed along ($\bar{2}$01) orientation. Results from this study on MOCVD growth and band offsets of amorphous $Al_2O_3$ deposited on differently oriented β-$Ga_2O_3$ and β-$(Al_xGa_{1-x})_2O_3$ films will potentially contribute to design and fabricate future high performance β-

$Ga_2O_3$ and β-$(Al_xGa_{1-x})_2O_3$ based transistors using MOCVD in-situ deposited $Al_2O_3$ as gate dielectric.

## I. Introduction

Beta gallium oxide (β-$Ga_2O_3$) with its ultrawide bandgap (UWBG) energy (4.5-4.9 eV) and high internal critical breakdown field strength ($E_{br,\ predicted}$ = 8 MV/cm) has stimulated substantial research interests as an emerging semiconductor material for next generation high power electronics [1]. The controllable n-type doping of β-$Ga_2O_3$ with excellent transport properties and its availability of differently oriented ((010), (100), $(\bar{2}01)$ and (001)) single crystal native substrates have paved the way for developing high performance lateral and vertical transistors, Schottky barrier diodes and ultraviolet photodetectors fabricated on high quality β-$Ga_2O_3$ epi-films [2-19]. Furthermore, alloying β-$Ga_2O_3$ with $Al_2O_3$ provides additional band gap engineering over a range of 4.87-8.82 eV [20], and thus a pathway towards enhanced mobility with better device performance based on β-$Ga_2O_3$/β-$(Al_xGa_{1-x})_2O_3$ heterostructures [21]. In addition to the successful demonstration of β-$Ga_2O_3$/β-$(Al_xGa_{1-x})_2O_3$ based modulation doped field effect transistors (MODFETs) [22-25], the higher predicted breakdown field strength (16 MV/cm for 80% Al content) [26] with controllable n-type doping of β-$(Al_xGa_{1-x})_2O_3$ [27-29] show a great potential for the development of high-power and high-frequency electronic devices surpassing the state-of-the-art.

In recent years, epitaxial development of β-$(Al_xGa_{1-x})_2O_3$ thin films [30-39] and the implementation of β-$(Al_xGa_{1-x})_2O_3$/β-$Ga_2O_3$ based MODFET devices [22-25] have made significant progresses. Metalorganic chemical vapor deposition (MOCVD) of β-$(Al_xGa_{1-x})_2O_3$ thin films on differently oriented β-$Ga_2O_3$ substrates such as (010) ($x \leq 0.35$) [28,30,34,35,38], (100) ($x \leq 0.52$) [31,32] and ($\bar{2}01$) ($x \leq 0.48$) [33,38] have been demonstrated. Molecular beam epitaxy (MBE) grown Si delta-doped MODFETs have been demonstrated with decent electron mobility (< 180 $cm^2$/V.s) and relatively high sheet charge density (5 x $10^{12}$ $cm^{-2}$) by using 18-20% Al composition in the β-$(Al_xGa_{1-x})_2O_3$ layer [22-24]. Recently, a room temperature Hall mobility of 111 $cm^2$/V.s with sheet charge density of 1.06 x $10^{13}$ $cm^{-2}$ was measured in MOCVD grown β-$Ga_2O_3$/β-$(Al_{0.27}Ga_{0.73})_2O_3$ modulation-doped heterostructures [25]. To achieve higher density two-dimensional electron gases (2DEGs), high Al incorporation in β-$(Al_xGa_{1-x})_2O_3$ layer with larger β-$Ga_2O_3$/β-$(Al_xGa_{1-x})_2O_3$ band offset is required. Apart from the different incorporation efficiency of Al composition in differently oriented β-$(Al_xGa_{1-x})_2O_3$ thin films [28,31,33], the experimentally measured band offsets at β-$Ga_2O_3$/β-$(Al_xGa_{1-x})_2O_3$ heterointerfaces also exhibited a strong dependence on the substrate orientation [38]. The (100) orientated β-$Ga_2O_3$/β-$(Al_xGa_{1-x})_2O_3$ heterointerfaces showed relatively higher conduction band offsets as compared to (010) and ($\bar{2}01$) orientations [32,38], indicating potential opportunities for excellent carrier confinement along (100) orientation.

While high quality and high-Al content β-$(Al_xGa_{1-x})_2O_3$ films with larger β-$Ga_2O_3$/β-$(Al_xGa_{1-x})_2O_3$ band offsets are advantageous for carrier confinement at the heterointerfaces, the quality of the dielectric-semiconductor interface is crucial for reliable high-performance transistors. Several gate dielectric materials such as $Al_2O_3$, $HfO_2$, $SiO_2$ and their alloys or bilayer combinations have been extensively investigated in β-$Ga_2O_3$ based metal oxide-semiconductor field effect transistors (MOSFETs) [40-45]. While high quality dielectric material with high dielectric constant and lower concentration of interface and bulk traps can significantly improve the performance of β-$Ga_2O_3$ and β-$(Al_xGa_{1-x})_2O_3$ based devices, the higher conduction band offset (> 1 eV) at dielectric/β-$Ga_2O_3$ (or β-$(Al_xGa_{1-x})_2O_3$) interfaces is also expected for developing MOS structures with large breakdown field and lower gate leakage. Due to its good compatibility with β-$Ga_2O_3$, $Al_2O_3$ with a dielectric constant of 7 - 8.5 has been widely used as a gate dielectric in β-$Ga_2O_3$ and β-$(Al_xGa_{1-x})_2O_3$ based devices. Excellent transistor performance in terms of high figure of merits with large breakdown field strength has been demonstrated in (010) and (001) oriented β-$Ga_2O_3$ based lateral and vertical FET structures using atomic layer deposited (ALD) $Al_2O_3$ as gate dielectric [15,46,47,48]. However, such ex-situ deposition of $Al_2O_3$ dielectrics using ALD or sputtering might result in interface contamination due to the exposure of the surface of β-$Ga_2O_3$ or β-$(Al_xGa_{1-x})_2O_3$ epi-films to the ambient during loading the samples into different chamber to deposit gate dielectrics. Instead, the in-situ deposition of gate dielectric can potentially avoid such surface or interface contamination of the epi-layer by avoiding the exposure to the ambient atmosphere. Recently, MOCVD in-situ $Al_2O_3$ deposition on β-$Ga_2O_3$ is demonstrated with reduced interface traps as compared to the commonly used ALD technique [49], indicating a great potential for depositing high quality gate dielectrics using MOCVD. Moreover, the high temperature deposition of MOCVD (450-1000 °C) with higher deposition rates can improve the quality of the bulk dielectric materials with reduced trap density at the interface. From previous studies, MOCVD in situ growth of amorphous $Al_2O_3$ on GaN has been demonstrated with lower trap densities (7.4 x $10^{11}$ $cm^{-2}eV^{-1}$) and very small hysteresis as compared to the ex-situ ALD $Al_2O_3$ (1.6 x $10^{12}$ $cm^{-2}eV^{-1}$) on GaN MOSCAPs [50]. One of the motivations of this study is to explore the feasibility of in-situ MOCVD growth of amorphous $Al_2O_3$ on β-$Ga_2O_3$ and β-$(Al_xGa_{1-x})_2O_3$, since majority of the dielectric $Al_2O_3$ on $Ga_2O_3$ based transistors was deposited by ex-situ ALD. However, further studies are still required to investigate the influence of MOCVD growth conditions on the device performance while using the in-situ $Al_2O_3$ as the gate dielectric in transistors. Indeed, MOCVD can provide a wide range of growth conditions that can lead to different interface properties between $Al_2O_3$ and $Ga_2O_3$ or $(Al_xGa_{1-x})_2O_3$. The prior report on ALD deposited $Al_2O_3$ on (010) β-$Ga_2O_3$ films revealed the formation of crystalline interlayers of γ-phase $Al_2O_3$ with 1-5 nm thickness at the interface of $Al_2O_3$ and β-$Ga_2O_3$ with lower interface state densities than those of the amorphous $Al_2O_3$/(-201) β-$Ga_2O_3$ interface [51], indicating the potential of achieving high quality interfaces due to the crystallization of amorphous $Al_2O_3$. While MOCVD growth of $Al_2O_3$ dielectric is demonstrated on both β-$Ga_2O_3$ [49] and GaN [50, 52] semiconductors, the investigation of the interface quality of in-situ MOCVD deposited $Al_2O_3$/β-$(Al_xGa_{1-x})_2O_3$ interfaces are still lacking. Considering the highly orientation dependent band offsets at β-$Ga_2O_3$/β-$(Al_xGa_{1-x})_2O_3$ interfaces [38], the investigation of the material quality of $Al_2O_3$ dielectric deposited on differently oriented β-$(Al_xGa_{1-x})_2O_3$ interfaces with different Al compositions are still in need. Moreover, the accurate extraction of the band offsets at dielectric/β-$(Al_xGa_{1-x})_2O_3$ interface is also critical for designing and fabricating β-$(Al_xGa_{1-x})_2O_3$ based devices. Nevertheless, information related to the band discontinuities at the interface of $Al_2O_3$ dielectric and differently oriented β-$(Al_xGa_{1-x})_2O_3$ as a function of Al composition are not reported yet.

In this work, we have systematically investigated the interface quality of in-situ MOCVD grown $Al_2O_3$ dielectric on (010), (100) and ($\bar{2}01$) oriented β-$Ga_2O_3$ and β-$(Al_xGa_{1-x})_2O_3$ thin films and experimentally determined the band offsets at $Al_2O_3$(dielectric)/β-$(Al_xGa_{1-x})_2O_3$ heterointerfaces with various Al compositions along different orientations. Moreover, the band offsets determined for MOCVD in-situ $Al_2O_3$ are also compared with ALD (ex-situ) $Al_2O_3$ deposited on β-$(Al_xGa_{1-x})_2O_3$ interfaces.

## II. Experimental Section

The β-$(Al_xGa_{1-x})_2O_3$ thin films were grown on Fe doped semi-insulating (010), (100), and ($\bar{2}$01) oriented β-$Ga_2O_3$ substrates (purchased from Novel Crystal Technology, Inc), with an unintentionally doped (UID) ~ 65 nm thick β-$Ga_2O_3$ as buffer layer via Agnitron Technology Agilis R&D MOCVD system. Triethylgallium (TEGa), Trimethylaluminum (TMAl), and pure $O_2$ were used as Ga, Al, and O precursors, respectively. Argon (Ar) was used as the carrier gas. The different Al compositions (x = 0 - 52%) of differently oriented β-$(Al_xGa_{1-x})_2O_3$ films were obtained by systematically tuning the [TMAl]/[TMAl+TEGa] molar flow ratio, growth temperature (880 °C), chamber pressure (20-80 torr) and group VI/III ratio [28,30,31,33]. To minimize potential contaminations from the growth surface, all substrates were cleaned ex-situ by using solvents and then in-situ by high temperature (920°C) annealing for 5 mins under $O_2$ atmosphere before the growths start. On top of the β-$(Al_xGa_{1-x})_2O_3$ thin films, a 40 nm thick $Al_2O_3$ dielectric layer was deposited in the same MOCVD chamber at a growth temperature of 700 °C and chamber pressure of 20 torr. The $O_2$ flow rate was fixed at 500 sccm. The $Al_2O_3$ deposition rate of 4 nm/min was obtained with TMAl molar flow rate of 5.46 μmol/min. For the band offset measurements by using XPS, three types of samples were prepared as shown in Figure S1 of the supplementary material: (a) 20-60 nm thick β-$(Al_xGa_{1-x})_2O_3$ films grown on UID β-$Ga_2O_3$ buffer layer on top of (010), (100) and ($\bar{2}$01) β-$Ga_2O_3$ substrates, (b) ~ 40 nm thick, and (c) thin (~2-5 nm) $Al_2O_3$ deposited on (010), (100) and ($\bar{2}$01) oriented β-$(Al_xGa_{1-x})_2O_3$ films with different Al compositions. The ALD deposition of $Al_2O_3$ was performed in a Picosun SUNALE R-150B by using Trimethylaluminum (TMAl) as the Al precursor and water ($H_2O$) as the $O_2$ precursor. For the band offset measurements, on top of the MOCVD grown differently oriented β-$(Al_xGa_{1-x})_2O_3$ films, $Al_2O_3$ layers with thick (~50 nm) and thin (~3 nm) thicknesses were deposited at 300 °C in the ALD chamber. Prior to the loading into the ALD chamber, all β-$(Al_xGa_{1-x})_2O_3$ samples were cleaned using acetone, IPA, and DI water.

High resolution XRD using a Bruker D8 Discover was performed to evaluate the quality and structures of $Al_2O_3$ dielectrics and β-$(Al_xGa_{1-x})_2O_3$ epi-films oriented along different directions. Surface morphologies were evaluated by atomic force microscopy (AFM) using Bruker AXS Dimension Icon. Surface stoichiometry, band gap energies and band offsets at $Al_2O_3$/β-$(Al_xGa_{1-x})_2O_3$ were determined using XPS measurements with an energy resolution of 0.1 eV. A Kratos Axis Ultra X-ray photoelectron spectrometer using monochromic Al Kα x-ray source with a photon energy of 1486.6 eV was used for the XPS measurement. C 1s peak (284.8 eV) from the adventitious carbon was used as a reference to calibrate the binding energy. As the bandgaps and band offsets at $Al_2O_3$/β-$(Al_xGa_{1-x})_2O_3$ were determined using the binding energy differences between the core level peak positions of any given sample, the correction of the peak positions does not impact the results. The electron pass energy was set at 20 eV during high resolution scanning and 80 eV for survey scanning. A Thermo Fisher Scientific Themis-Z scanning transmission electron microscope (200 kV, Cs3=0.002 mm, and Cs5=1.0 mm) with probe convergence half angles of 17.9 mrad was utilized to perform high resolution STEM high angle annular dark filed (HAADF) imaging.

## III. Results and Discussions

To evaluate the crystalline phase and quality of the in-situ MOCVD deposited $Al_2O_3$ dielectrics grown on differently oriented β-$Ga_2O_3$ thin films, high resolution XRD was performed. Figures 1 (a)-(c) show the XRD ω-2θ scans of the ~ 40 nm thick $Al_2O_3$ dielectrics deposited on top of 65 nm thick UID β-$Ga_2O_3$ buffer layer on (100), ($\bar{2}01$) and (010) oriented β-$Ga_2O_3$ substrates, respectively. The $Al_2O_3$ dielectrics were deposited at a growth temperature of 700 °C. While high intensity distinguishable peaks originated from differently oriented β-$Ga_2O_3$ substrates are observed for all orientations, no other well-resolved XRD peaks corresponding to different crystalline phases/orientations of $Al_2O_3$ films are visible in wide range scans, indicating that the MOCVD deposited $Al_2O_3$ films possess amorphous structures for all different orientations of substrates, which is also evidenced from STEM imaging as discussed in the later sections. Our previous studies demonstrated the growth of γ-phase $Al_2O_3$ thin films on (010) β-$Ga_2O_3$ substrate at relatively higher growth temperature (880 °C) [30], suggesting that the lower growth temperature (700 °C) is a key parameter that can lead to the deposition of amorphous $Al_2O_3$ dielectrics on β-$Ga_2O_3$. The surface RMS roughness of a gate dielectric is also a critical parameter that has a significant impact on device performance. Figures 2 (a)-(c) show the surface RMS roughness of ~ 40 nm thick $Al_2O_3$ dielectric deposited in MOCVD chamber on (010), (100) and ($\bar{2}$01) β-$Ga_2O_3$ films, respectively. Smooth and uniform surface morphologies of $Al_2O_3$ with RMS roughness of 1.42 nm (010), 1.25 nm (100), and 1.17 nm ($\bar{2}$01) are observed on differently oriented β-$Ga_2O_3$ films. The lowest surface morphologies observed on ($\bar{2}$01) β-$Ga_2O_3$ films as compared to those of (010) and (100) β-$Ga_2O_3$, indicating a strong dependence of the surface morphologies of top $Al_2O_3$ dielectrics on the orientations of β-$Ga_2O_3$ substrates.

The quality of $Al_2O_3$ dielectrics and the interfacial abruptness of MOCVD deposited $Al_2O_3$/β-$Ga_2O_3$ (β-$(Al_xGa_{1-x})_2O_3$) for different orientations are investigated using high-resolution STEM imaging and energy dispersive x-ray spectroscopy (STEM-EDS) elemental mapping. High resolution HAADF STEM images for $Al_2O_3$/(010) β-$Ga_2O_3$ and $Al_2O_3$/(010) β-$(Al_{0.17}Ga_{0.83})_2O_3$ are shown in Figures 3 (a) and (b), respectively. The high resolution HAADF-STEM images of $Al_2O_3$ dielectric deposited on ($\bar{2}$01) and (100) oriented β-$Ga_2O_3$ and β-$(Al_{0.17}Ga_{0.83})_2O_3$ thin films are also shown in Figures 3 (c) and (d) [($\bar{2}$01) orientation] and (e) and (f) [(100) orientation], respectively. The growth of amorphous $Al_2O_3$ dielectrics on top of β-$Ga_2O_3$ and β-$(Al_{0.17}Ga_{0.83})_2O_3$ layers is confirmed from the atomic resolution STEM images. The sharp contrasts between $Al_2O_3$ dielectric (dark) and β-$Ga_2O_3$ (β-$(Al_{0.17}Ga_{0.83})_2O_3$) (bright) indicate high quality interfaces along all investigated orientations.

STEM-EDS spectroscopy also reveals abrupt and sharp interfaces between $Al_2O_3$ and (010) oriented β-$Ga_2O_3$ and β-$(Al_{0.17}Ga_{0.83})_2O_3$ as shown in Figures 4 (a-h) and Figures S2 and S3 of the supplementary materials. The EDS color maps of Ga (green) and Al (blue) elements in Figures 4 (a-d) [$Al_2O_3$/β-$Ga_2O_3$] and (e-h) [$Al_2O_3$/β-$(Al_{0.17}Ga_{0.83})_2O_3$] and elemental mapping (Figs. 4 (d) and (h)) also confirm the interfacial abruptness between $Al_2O_3$ and (010) oriented β-$Ga_2O_3$ (β-$(Al_{0.17}Ga_{0.83})_2O_3$). Figures 5 (a-d) and (e-h) show the HAADF-STEM images of $Al_2O_3$/($\bar{2}$01) β-$Ga_2O_3$ (Fig. 5(a)) and $Al_2O_3$/($\bar{2}$01) β-$(Al_{0.17}Ga_{0.83})_2O_3$ (Fig. 5(e)) interfaces with corresponding EDS maps of (b, f) Al, (c, g) Ga and (d, h) O elements. The abrupt transition of Ga and Al at $Al_2O_3$/($\bar{2}$01) β-$Ga_2O_3$ (β-$(Al_{0.17}Ga_{0.83})_2O_3$) interfaces are observed for both samples, as also shown in Figures S4 and S5 of the supplementary materials, with a very slight interdiffusion of Ga and Al atoms at the interfaces. The abruptness of the interfaces between $Al_2O_3$ and (100) oriented β-$Ga_2O_3$ (β-$(Al_xGa_{1-x})_2O_3$) is also evaluated by cross-sectional HAADF-STEM images and EDS elemental mapping of $Al_2O_3$/β-$Ga_2O_3$ and $Al_2O_3$/β-$(Al_{0.17}Ga_{0.83})_2O_3$ as shown in Figures 6 (a-d) and (e-h), (Figs. S6 and S7 of the supplementary materials) respectively. The EDS color maps of Al (blue) and Ga (green) components in Figures 6 (b, c) and (f, g) correspond to the Ga and Al concentrations of the samples as indicated by the EDS quantitative line scan (Figs. 6(d, h)) along the orange arrows in Figures 6 (a, e) reveal sharp and high-quality interfaces between $Al_2O_3$ dielectric and (100) oriented β-$Ga_2O_3$ (β-$(Al_{0.17}Ga_{0.83})_2O_3$) layers.

The surface stoichiometry of in-situ MOCVD deposited $Al_2O_3$ dielectric are investigated by using high resolution XPS. The O 1s, Al 2s and 2p core level data were used to determine the O and Al percentage in $Al_2O_3$ dielectric. Figures 7 (a-c) show the O 1s, Al 2s and Al 2p spectra for a representative $Al_2O_3$ film deposited on (010) β-$Ga_2O_3$. The peaks are fitted with Gaussian-Lorentzian function after subtracting by Shirley type background. The O 1s peak is fitted with two components, O-Al bond at 531.1 eV and O-H bond at 532.1 eV, as shown in Figure 7(a). The Al 2s and Al 2p core level peaks are fitted with single peak as shown in Figures 7(b) and (c), respectively. Using the relative sensitivity factors of O 1s, Al 2s and Al 2p peaks ($S_{O\ 1s} = 2.93$, $S_{Al\ 2s} = 0.753$ and $S_{Al\ 2p} = 0.5371$), the percentage of Al and O in $Al_2O_3$ dielectric is determined to be 40.39% and 59.61%, respectively. The list of all Al and O percentages in both MOCVD and ALD deposited $Al_2O_3$ on other orientations ((($\bar{2}01$)) and (100)) of β-$Ga_2O_3$ are included in Table S1 of the supplementary materials, indicating that the surface stoichiometries of $Al_2O_3$ deposited by using both MOCVD and ALD methods are well maintained on all investigated orientations.

By utilizing XPS, the bandgap of MOCVD deposited $Al_2O_3$ is estimated by measuring the onset of inelastic loss spectrum of O 1s core level peak. The estimation of the bandgap energies of wide bandgap semiconductor materials using the inelastic loss curve appeared on higher binding energy side of a high intensity XPS core level spectra is a well-established method [53-61]. The lower limit of inelastic scattering represents the bandgap of a material as the lowest energetically inelastic scattering that an electron experiences during its travelling from the bulk to the surface is the excitation from the valence band to the conduction band. Figure 8(a) shows the O 1s core level peak for a representative $Al_2O_3$ sample grown on (010) oriented β-$Ga_2O_3$ film. The onset is determined from the intersection of the linear extrapolation of the loss spectra curve and constant background as shown in Figure 8(b). The bandgap of MOCVD deposited $Al_2O_3$ is estimated as 6.91 eV from the separation between the onset and O 1s peak positions. Similarly, the ALD deposited $Al_2O_3$ also exhibited a bandgap energy of 6.88 eV, which is well consistent with the bandgap energy reported previously from different deposition methods [55-57, 62-64]. The bandgap determined by Reflection Electron Energy Loss Spectroscopy were estimated as 6.9 eV for $Al_2O_3$ deposited by either ALD or sputtering method [62-64], which are found to be very consistent with our extracted bandgap values by using XPS. Apart from these studies, the band gap of amorphous $Al_2O_3$ was determined to be 7.0 ± 0.1 eV by using Al 2p and O 1s photoelectron spectra [55]. In addition, 6.8 eV of bandgap energy were measured using O 1s core-level binding energy spectrum of ALD-$Al_2O_3$ on (010) $Ga_2O_3$ [56]. Similarly, from the energy separation between the main O 1s line and the loss threshold, a bandgap of 7.0 ± 0.1 eV was determined for the 120 nm thick $Al_2O_3$ film deposited on 6H-SiC [57]. These studies indicate that our measured bandgap of both MOCVD and ALD deposited $Al_2O_3$ are in a good agreement with previous reports. Using the same approach, the extraction of bandgap energies of (010), (100) and ($\bar{2}01$) oriented β-$Ga_2O_3$ and β-$(Al_xGa_{1-x})_2O_3$ with different Al compositions [38] are listed in Table S2 of the supplementary material.

The band offsets at the interfaces of $Al_2O_3$ with all three orientations ((010), (100) and ($\bar{2}01$)) of β-$Ga_2O_3$ and β-$(Al_xGa_{1-x})_2O_3$ for a wide Al composition range are determined by using XPS. The valence band offsets ($\Delta E_v$) at $Al_2O_3$/β-$Ga_2O_3$ (β-$(Al_xGa_{1-x})_2O_3$) are determined by using the Kraut's method [65]. Using the bandgaps of $Al_2O_3$, β-$Ga_2O_3$ and β-$(Al_xGa_{1-x})_2O_3$ and $\Delta E_v$ values, the corresponding conduction band offsets ($\Delta E_c$) are estimated as follows:

$$\Delta E_v = (E^{AlGaO}_{Ga\ 3s} - E^{AlGaO}_{VBM}) - (E^{AlO}_{Al\ 2p} - E^{AlO}_{VBM}) - (E^{AlGaO/AlO}_{Ga\ 3s} - E^{AlGaO/AlO}_{Al\ 2p}) \quad (1)$$

$$\Delta E_c = E^{AlO}_{g} - E^{AlGaO}_{g} - \Delta E_v \quad (2)$$

where $E_{\mathrm{Ga\ 3s}}^{\mathrm{AlGaO}}$ and $E_{Ga\ 3s}^{\mathrm{AlGaO/AlO}}$ are the binding energies of Ga 3s core levels in β-$(Al_xGa_{1-x})_2O_3$ bulk material and $Al_2O_3$/β-$(Al_xGa_{1-x})_2O_3$ interfaces, respectively. $E_{\mathrm{Al\ 2p}}^{\mathrm{AlO}}$ and $E_{\mathrm{Al\ 2p}}^{\mathrm{AlGaO/AlO}}$ are the Al 2p core level binding energies of $Al_2O_3$ bulk material and $Al_2O_3$/β-$(Al_xGa_{1.x})_2O_3$ interfaces, respectively. $E_{\mathrm{VBM}}^{\mathrm{AlGaO}}$ and $E_{\mathrm{VBM}}^{\mathrm{AlO}}$ are the valence band maxima (VBM) positions of β-$(Al_xGa_{1-x})_2O_3$ and $Al_2O_3$ bulk materials, respectively. $E_{\mathrm{g}}^{\mathrm{AlO}}$ and $E_{\mathrm{g}}^{\mathrm{AlGaO}}$ are the bandgap energies of the $Al_2O_3$ and β-$(Al_xGa_{1-x})_2O_3$ thin films, respectively.

The fitted Ga 3s, Al 2p core level peaks and the valence band (VB) spectra are shown in Figures 9(a)-(f) for the in-situ MOCVD deposited $Al_2O_3$ on (010) oriented β-$(Al_{0.17}Ga_{0.83})_2O_3$ thin films. The Ga 3s (Figs. 9(a) and (e)) and Al 2p (Figs. 9(c) and (f)) core level peak positions for bulk β-$(Al_{0.17}Ga_{0.83})_2O_3$ (Figs. 9(a)), $Al_2O_3$ (Figs. 9(c)) and $Al_2O_3$/β-$(Al_{0.17}Ga_{0.83})_2O_3$ heterostructure (Figs. 9(e) and (f)) are determined by fitting with mixed Lorentzian-Gaussian line shapes after applying the Shirley type background subtraction. Both Ga 3s and Al 2p core levels are fitted with the single peak. The VBM positions from bulk β-$(Al_{0.17}Ga_{0.83})_2O_3$ (Figs. 9(b)) and $Al_2O_3$ (Figs. 9(d)) are determined by the linear extrapolation of the leading edge of the VB spectra to the background. The $\Delta E_v$ of -0.18 eV is determined at the $Al_2O_3$/(010) β-$(Al_{0.17}Ga_{0.83})_2O_3$ interface by using equation (1). By using the $\Delta E_v$ value and bandgaps of $Al_2O_3$ and (010) β-$(Al_{0.17}Ga_{0.83})_2O_3$, the $\Delta E_c$ of 2.06 eV is determined from equation (2). Using similar approach, the band offsets at the interfaces between MOCVD (and ALD) deposited $Al_2O_3$ dielectrics and differently oriented β-$Ga_2O_3$ and β-$(Al_xGa_{1-x})_2O_3$ heterostructures for all investigated Al compositions are determined. The band offsets determined by using Ga 3s and Al 2p core levels are also verified by comparing with those extracted from Ga 3s /Ga 3d and Al 2p / Al 2s core levels as listed exemplarily for $Al_2O_3$/(010) β-$(Al_xGa_{1-x})_2O_3$ interfaces in Table 1 and Tables S2 and S3 of the supplementary material for other $Al_2O_3$/β-$(Al_xGa_{1-x})_2O_3$ interfaces along different orientations and deposition methods, indicating a good consistency between the band offsets values estimated from different core level spectra.

Heterostructures are classified as Type I and Type II depending on the alignment of the bands producing the discontinuity. While one material's band gap overlaps another's in Type I- straddling heterostructures, both conduction and valence band edges of a material are lower than the corresponding band edges of another material in Type II- staggered heterostructure. The band alignment diagram of MOCVD deposited $Al_2O_3$/β-$Ga_2O_3$ (β-$(Al_xGa_{1-x})_2O_3$) heterostructures are shown in Figures 10 (a)-(c) for (010), (100) and ($\bar{2}01$) orientations, respectively. Type II (staggered) band alignment at the interfaces of $Al_2O_3$ deposited on (010) and (100) oriented β-$(Al_xGa_{1-x})_2O_3$ is determined for all investigated Al compositions, whereas the ($\bar{2}01$) orientations show type I (straddling) band alignment between $Al_2O_3$ and β-$(Al_xGa_{1-x})_2O_3$. The valence band offsets are found to vary from -0.07 to -0.17 eV (010), -0.14 to -0.23 eV (100) and 0.20 to 0.27 eV ($\bar{2}01$) with corresponding conduction band offsets ranging between 2.14 and 1.66 eV (010), 2.22 and 1.29 eV (100), and 1.79 to 0.9 eV ($\bar{2}01$) as the Al compositions in differently oriented β-$(Al_xGa_{1-x})_2O_3$ are varied from 0 to 35% (010), 0 to 52% (100) and 0 to 48% ($\bar{2}01$). Similarly, the evolution of the band alignments at the interfaces of ALD deposited $Al_2O_3$ with the MOCVD grown β-$(Al_xGa_{1-x})_2O_3$ with different Al compositions are shown in Figures 11 (a)-(c) for (010), (100) and ($\bar{2}01$) orientations, respectively. As the Al composition varies, the valence and conduction band offsets range from -0.12 to -0.35 eV ($\Delta E_v$) and 2.18 to 1.81 eV ($\Delta E_c$) (for (010) orientation), -0.31 to -0.55 eV ($\Delta E_v$) and 2.36 to 1.44 eV ($\Delta E_c$) (for (100) orientation) and 0.07 to 0.29 eV ($\Delta E_v$) and 1.74 to 0.89 eV ($\Delta E_c$) (for ($\bar{2}01$) orientation), respectively. Similar to the in-situ MOCVD deposited $Al_2O_3$, type II band alignments are observed for both (010) and (100) orientations and type I band discontinuity is observed at the interface of ALD $Al_2O_3$/($\bar{2}01$) β-$(Al_xGa_{1-x})_2O_3$. For both MOCVD and ALD deposition of $Al_2O_3$, the conduction band offsets exhibit relatively larger variations with Al compositions as compared to the valence band offsets for all three orientations. Such lower $\Delta E_v$ (0.25 eV) as compared to the conduction band offset ($\Delta E_c$) of 2.06 eV was also demonstrated between the interface of amorphous $Al_2O_3$ and monoclinic β-$Ga_2O_3$ [66] by using the band-edge position information of amorphous $Al_2O_3$ from experimentally determined band gap and valence-band offset with corundum α-phase $Al_2O_3$ [55].

The evolution of the valence and conduction band offsets of both in-situ MOCVD and ALD grown $Al_2O_3$ dielectrics on β-$(Al_xGa_{1-x})_2O_3$ heterointerfaces are represented in Figures 12 (a)-(d) as a function of Al composition for different orientations. The general trend shows that both $\Delta E_v$ and $\Delta E_c$ values decrease as the Al compositions in β-$(Al_xGa_{1-x})_2O_3$ increases, with a much weaker variation in the $\Delta E_v$ values as compared to the corresponding changes in $\Delta E_c$ values. Although both deposition methods showed type II band discontinuity along (010) and (100) orientations and type I band alignment along ($\bar{2}01$) orientations, the variations in the values of $\Delta E_v$ and $\Delta E_c$ between MOCVD and ALD deposited $Al_2O_3$/β-$(Al_xGa_{1-x})_2O_3$ heterointerfaces can be attributed to the quality of the bulk $Al_2O_3$ and $Al_2O_3$/β-$(Al_xGa_{1-x})_2O_3$ interfaces as the deposition of $Al_2O_3$ is performed under completely different growth environments. Previously, type I straddling gap band alignment was demonstrated on ALD $Al_2O_3$ on ($\bar{2}01$) β-$Ga_2O_3$, whereas sputtered $Al_2O_3$ showed a type II staggered gap with β-$Ga_2O_3$ [62]. Similar changes of band alignment from type I to type II are also observed as the $Al_2O_3$ deposition methods changed from ALD to sputtering on (010) β-$(Al_{0.14}Ga_{0.86})_2O_3$ interfaces [64], revealing a strong influence of the synthesis method on the resulting band alignment. In this work, for both MOCVD and ALD deposition, the $Al_2O_3$/($\bar{2}01$) β-$(Al_xGa_{1-x})_2O_3$ interfaces show the smallest conduction band offsets as compared to the (010) and (100) orientations, with a type I band alignment (Figs. 12(b,d)). While a small variation in $\Delta E_v$ (Fig. 12(a)) and $\Delta E_c$ (Fig. 12(b)) values between the (010) and (100) orientations are observed for in-situ MOCVD deposited $Al_2O_3$/β-$(Al_xGa_{1-x})_2O_3$ interfaces, the ALD deposited $Al_2O_3$ on (100) oriented β-$(Al_xGa_{1-x})_2O_3$ exhibits relatively higher $\Delta E_v$ (Fig. 12(c)) and $\Delta E_c$ (Fig. 12(d)) values as compared to other orientations, indicating a strong dependence of the band offsets on dielectric deposition methods, crystalline orientations and Al compositions of β-$(Al_xGa_{1-x})_2O_3$ films.

## IV. Conclusion

In summary, in-situ MOCVD growth of $Al_2O_3$ dielectrics with high quality $Al_2O_3$/β-$(Al_xGa_{1-x})_2O_3$ interfaces are successfully demonstrated on different orientations of β-$(Al_xGa_{1-x})_2O_3$ as a function of Al composition. The growth of high quality amorphous $Al_2O_3$ dielectrics with sharp interfaces and good surface stoichiometry are revealed by comprehensive characterization of high resolution XRD, STEM and XPS. As compared to (010) and (100) orientations, relatively lower RMS roughness is observed for $Al_2O_3$ dielectrics deposited on $(\bar{2}01)$ β-$Ga_2O_3$, indicating an influence of crystal orientation of underlying β-$Ga_2O_3$ on the surface morphologies of top $Al_2O_3$ dielectric. The band offsets between in-situ MOCVD and ALD deposited $Al_2O_3$ dielectric and β-$(Al_xGa_{1-x})_2O_3$ interfaces are investigated systematically for a wide range of Al compositions, revealed a strong influence of deposition methods, orientations, and Al composition of β-$(Al_xGa_{1-x})_2O_3$ films on the valence and conduction band offsets. A type-II band alignment at $Al_2O_3$/β-$(Al_xGa_{1-x})_2O_3$ interfaces for all investigated Al compositions are observed for (010) and (100) oriented β-$(Al_xGa_{1-x})_2O_3$ films, whereas type-I band alignment is observed for $(\bar{2}01)$ orientation with relatively lower $\Delta E_c$ values for both deposition methods. A weaker variation in $\Delta E_v$ values with different Al compositions of β-$(Al_xGa_{1-x})_2O_3$ are observed as compared to the variation in $\Delta E_c$ values. The high quality MOCVD growth of $Al_2O_3$ on differently oriented β-$Ga_2O_3$ and β-$(Al_xGa_{1-x})_2O_3$ films with sharp interfaces and the information on orientation dependent band offsets will be useful for the understanding, modeling and fabrication of β-$Ga_2O_3$ and β-$(Al_xGa_{1-x})_2O_3$ based MOS devices.

See the supplementary material for the schematic of the samples grown for band offset measurements, summary of the O and Al percentage of $Al_2O_3$ dielectrics calculated by using O 1s and Al 2p/Al 2s core level XPS spectra, and the summary of the band offsets at MOCVD in-situ and ALD ex-situ deposited $Al_2O_3$/β-$(Al_xGa_{1-x})_2O_3$ interfaces.

**Conflict of Interest Statement**

On behalf of all authors, the corresponding author states that there is no conflict of interest.

**Supplementary Material**

See supplementary material for XPS data for determining the band offsets and STEM-EDS maps.

**Acknowledgements**

The authors acknowledge the Air Force Office of Scientific Research FA9550-18-1-0479 (AFOSR, Dr. Ali Sayir) for financial support. The authors also acknowledge the National Science Foundation (Grant No. 1810041, No. 2019753) and Semiconductor Research Corporation (SRC) under the Task ID GRC 3007.001 for partial support. Electron microscopy was performed at the Center for Electron Microscopy and Analysis (CEMAS) at The Ohio State University.

**Data Availability**

The data that support the findings of this study are available from the corresponding author upon reasonable request.

## Table Caption

**Table 1.** Summary of the valence and conduction band offsets at $Al_2O_3$/(010) β-$(Al_xGa_{1-x})_2O_3$ interfaces, determined by using the VBM, Ga 3s, Ga 3d and Al 2p core level peak positions.

## Figure Captions

**Figure 1.** XRD ω-2θ scans of in-situ MOCVD grown ~ 40 nm thick $Al_2O_3$ dielectrics on (a) (010), (b) $(\bar{2}01)$, and (c) (100) β-$Ga_2O_3$ substrates.

**Figure 2.** Surface AFM images (scan area: 5x5 $\mu m^2$) of MOCVD deposited $Al_2O_3$ dielectrics on (a) (010), (b) $(\bar{2}01)$, and (c) (100) β-$Ga_2O_3$ substrates.

**Figure 3.** High resolution HAADF-STEM cross-sectional images of in-situ MOCVD deposited $Al_2O_3$ dielectrics grown on (a) (010) β-$Ga_2O_3$, (b) (010) β-$(Al_{0.17}Ga_{0.83})_2O_3$, (c) $(\bar{2}01)$ β-$Ga_2O_3$, (d) $(\bar{2}01)$ β-$(Al_{0.17}Ga_{0.83})_2O_3$, (e) (100) β-$Ga_2O_3$, and (f) (100) β-$(Al_{0.17}Ga_{0.83})_2O_3$ layers. The HAADF-STEM images for $Al_2O_3$/(010) β-$Ga_2O_3$ (β-$(Al_{0.17}Ga_{0.83})_2O_3$) are taken from [001] zone axis and $Al_2O_3$/$(\bar{2}01)$ and (100) oriented β-$Ga_2O_3$ (β-$(Al_{0.17}Ga_{0.83})_2O_3$) are taken from [010] zone axis.

**Figure 4.** STEM-EDS maps for MOCVD deposited $Al_2O_3$ dielectrics on (a-d) (010) β-$Ga_2O_3$ and (e-h) (010) β-$(Al_{0.17}Ga_{0.83})_2O_3$. HAADF images of (a) $Al_2O_3$/(010) β-$Ga_2O_3$ and $Al_2O_3$/(010) β-$(Al_{0.17}Ga_{0.83})_2O_3$ with corresponding (b,f) Al and (c,g ) Ga EDS maps and (d,h) atomic fraction elemental profiles as indicated by the orange arrow in (a, e).

**Figure 5.** STEM-EDS for maps MOCVD $Al_2O_3$ dielectrics deposited on (a-d) $(\bar{2}01)$ β-$Ga_2O_3$ and (e-h) $(\bar{2}01)$ β-$(Al_{0.17}Ga_{0.83})_2O_3$. HAADF images of (a) $Al_2O_3$/$(\bar{2}01)$ β-$Ga_2O_3$ and $Al_2O_3$/$(\bar{2}01)$ β-$(Al_{0.17}Ga_{0.83})_2O_3$ with corresponding (b,f) Al and (c,g ) Ga EDS maps and (d,h) atomic fraction elemental profiles as indicated by the orange arrow in (a, e).

**Figure 6.** STEM-EDS maps for MOCVD $Al_2O_3$ dielectrics deposited on (a-d) (100) β-$Ga_2O_3$ and (e-h) (100) β-$(Al_{0.17}Ga_{0.83})_2O_3$. HAADF images of (a) $Al_2O_3$/(100) β-$Ga_2O_3$ and $Al_2O_3$/(100) β-

$(Al_{0.17}Ga_{0.83})_2O_3$ with corresponding (b,f) Al and (c,g ) Ga EDS maps and (d,h) atomic fraction elemental profiles as indicated by the orange arrow in (a, e).

**Figure 7.** XPS (a) O 1s, (b) Al 2s and (c) Al 2p core level spectra for in-situ MOCVD grown $Al_2O_3$ dielectrics on (010) $\beta$-$Ga_2O_3$ substrates. Experimental data points (black open circles) are fitted using mixed Lorentzian-Gaussian line shapes (black solid lines) after applying the Shirley background (gray solid lines) subtraction.

**Figure 8.** (a) The bandgap energy estimated by using the onset of inelastic loss spectra of O 1s core level peak from in-situ MOCVD grown $Al_2O_3$ dielectrics on (010) $\beta$-$(Al_{0.17}Ga_{0.83})_2O_3$. (b) zoomed view of the background region of the O 1s core level.

**Figure 9.** Fitted (a) Ga 3s and (b) valence band (VB) spectra from (010) $\beta$-$(Al_{0.17}Ga_{0.83})_2O_3$ and (c) Al 2p and (d) VB spectra from ~40 nm thick $Al_2O_3$ deposited in-situ by MOCVD. The fitted Ga 3s and Al 2p core levels from $Al_2O_3$/(010) $\beta$-$(Al_{0.17}Ga_{0.83})_2O_3$ heterointerfaces are shown in (e) and (f), respectively. Experimental data are shown as black open circles and the fitted curves (envelope) are represented as black sloid lines.

**Figure 10.** The band alignments at the interfaces of in-situ MOCVD deposited $Al_2O_3$ and (a) (010) $\beta$-$(Al_xGa_{1-x})_2O_3$ (x = 0 - 0.35) (b) (100) $\beta$-$(Al_xGa_{1-x})_2O_3$ (x = 0 - 0.52) and (c) $(\bar{2}01)$ oriented $\beta$-$(Al_xGa_{1-x})_2O_3$ (x = 0 - 0.48).

**Figure 11.** The band alignments at the interfaces of ALD deposited $Al_2O_3$ and (a) (010) $\beta$-$(Al_xGa_{1-x})_2O_3$ (x = 0 - 0.35) (b) (100) $\beta$-$(Al_xGa_{1-x})_2O_3$ (x = 0 - 0.52) and (c) $(\bar{2}01)$ oriented $\beta$-$(Al_xGa_{1-x})_2O_3$ (x = 0 - 0.48).

**Figure 12.** The evolution of the (a,c) valence band offsets (considering the valence band maxima position of $Al_2O_3$ at 0 eV) and (b,d) conduction band offsets at the interfaces of (a,b) in-situ

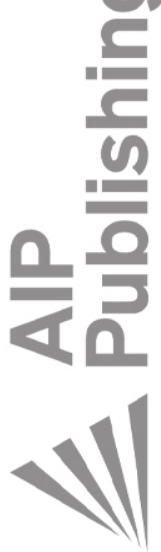

MOCVD and (c,d) ALD deposited $Al_2O_3$ and differently oriented $\beta$-$(Al_xGa_{1-x})_2O_3$ as a function of Al compositions.

**Table 1.**

**Ga 3s and Al 2p:**

| **Sample** | **$E_g$ (± 0.20 eV)** | **$E^{AlGaO}_{Ga\ 3s}$ - $E^{AlGaO}_{VBM}$ (± 0.04 eV)** | **$E^{AlO}_{Al\ 2p}$ - $E^{AlO}_{VBM}$ (± 0.04 eV)** | **$E^{AlGaO/AlO}_{Ga\ 3s}$ - $E^{AlGaO/AlO}_{Al\ 2p}$ (± 0.02 eV)** | **$\Delta E_v$ (eV) (± 0.06 eV)** | **$\Delta E_c$ (eV) (± 0.29 eV)** |
|---|---|---|---|---|---|---|
| $Al_2O_3$ | 6.91 | | 71.23 | | | |
| β-$Ga_2O_3$ | 4.84 | 157.43 | | 86.27 | -0.07 | 2.14 |
| β-$(Al_{0.17}Ga_{0.83})_2O_3$ | 5.03 | 157.38 | | 86.32 | -0.18 | 2.06 |
| β-$(Al_{0.35}Ga_{0.65})_2O_3$ | 5.42 | 157.47 | | 86.40 | -0.17 | 1.66 |

**Ga 3d and Al 2p:**

| **Sample** | **$E_g$ (± 0.20 eV)** | **$E^{AlGaO}_{Ga\ 3d}$ - $E^{AlGaO}_{VBM}$ (± 0.04 eV)** | **$E^{AlO}_{Al\ 2p}$ - $E^{AlO}_{VBM}$ (± 0.04 eV)** | **$E^{AlGaO/AlO}_{Ga\ 3d}$ - $E^{AlGaO/AlO}_{Al\ 2p}$ (± 0.02 eV)** | **$\Delta E_v$ (eV) (± 0.06 eV)** | **$\Delta E_c$ (eV) (± 0.29 eV)** |
|---|---|---|---|---|---|---|
| $Al_2O_3$ | 6.91 | | 71.23 | | | |
| β-$Ga_2O_3$ | 4.84 | 16.87 | | -54.32 | -0.05 | 2.12 |
| β-$(Al_{0.17}Ga_{0.83})_2O_3$ | 5.03 | 16.79 | | -54.30 | -0.14 | 2.02 |
| β-$(Al_{0.35}Ga_{0.65})_2O_3$ | 5.42 | 16.87 | | -54.24 | -0.12 | 1.61 |

**Figure 1.**

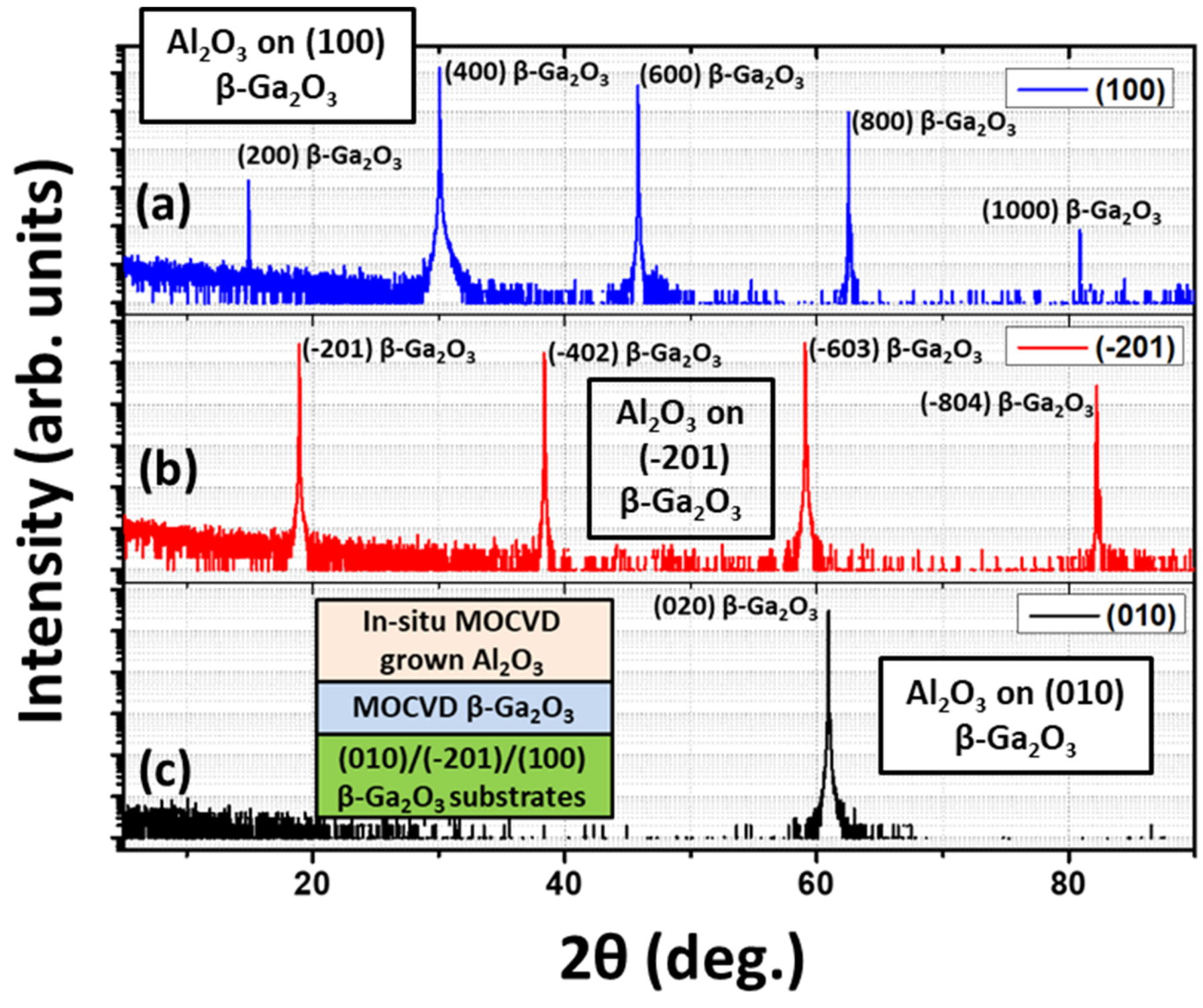

**Figure 2.**

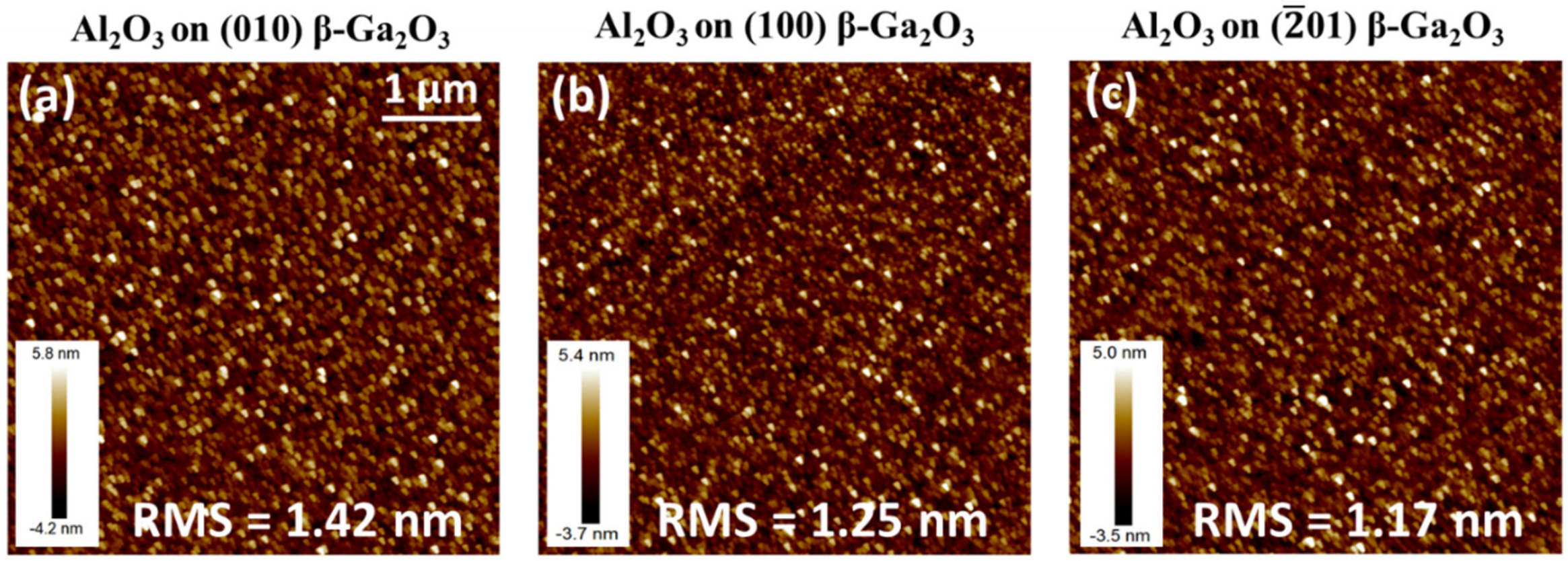

**Figure 3.**

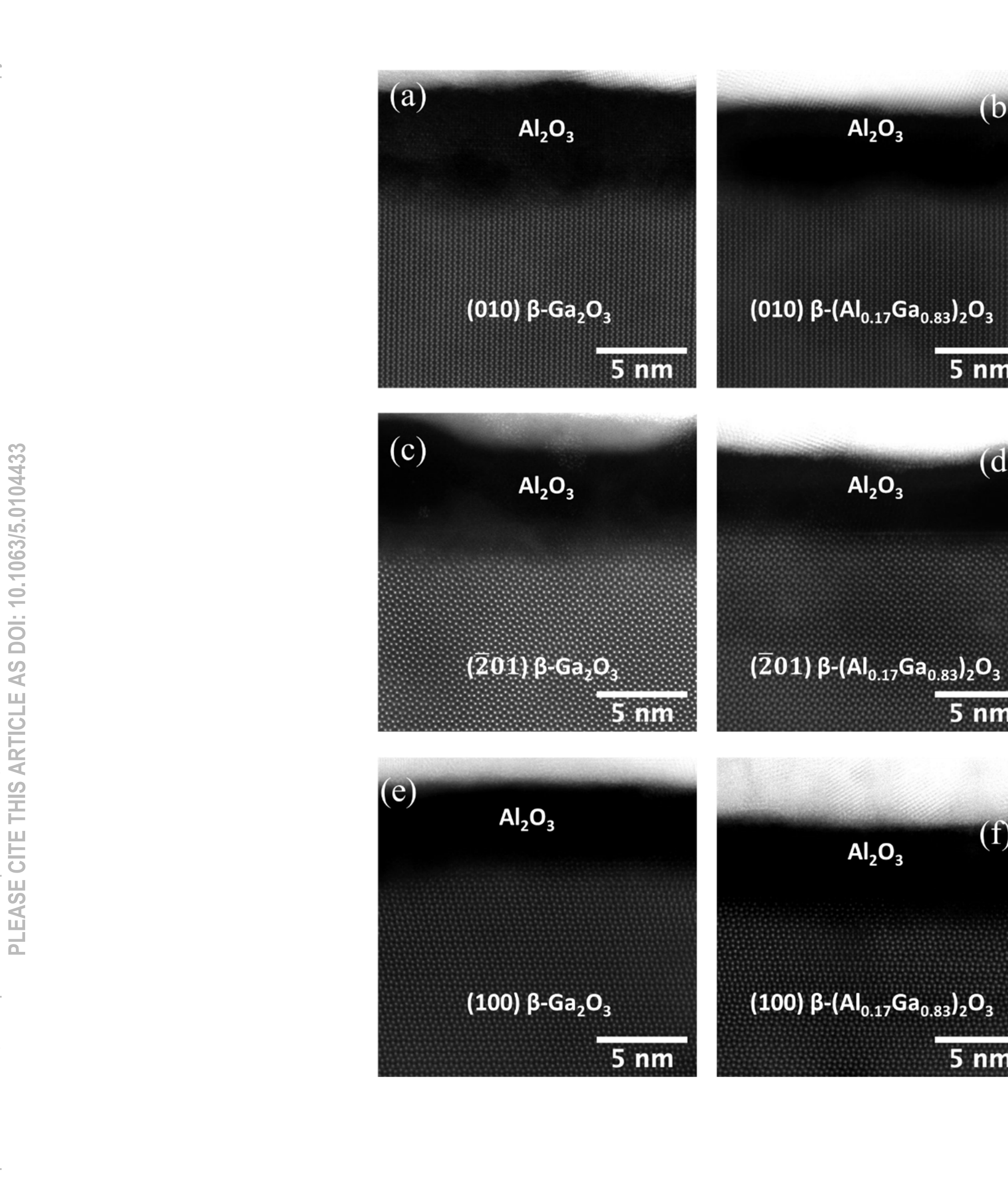

**Figure 4.**

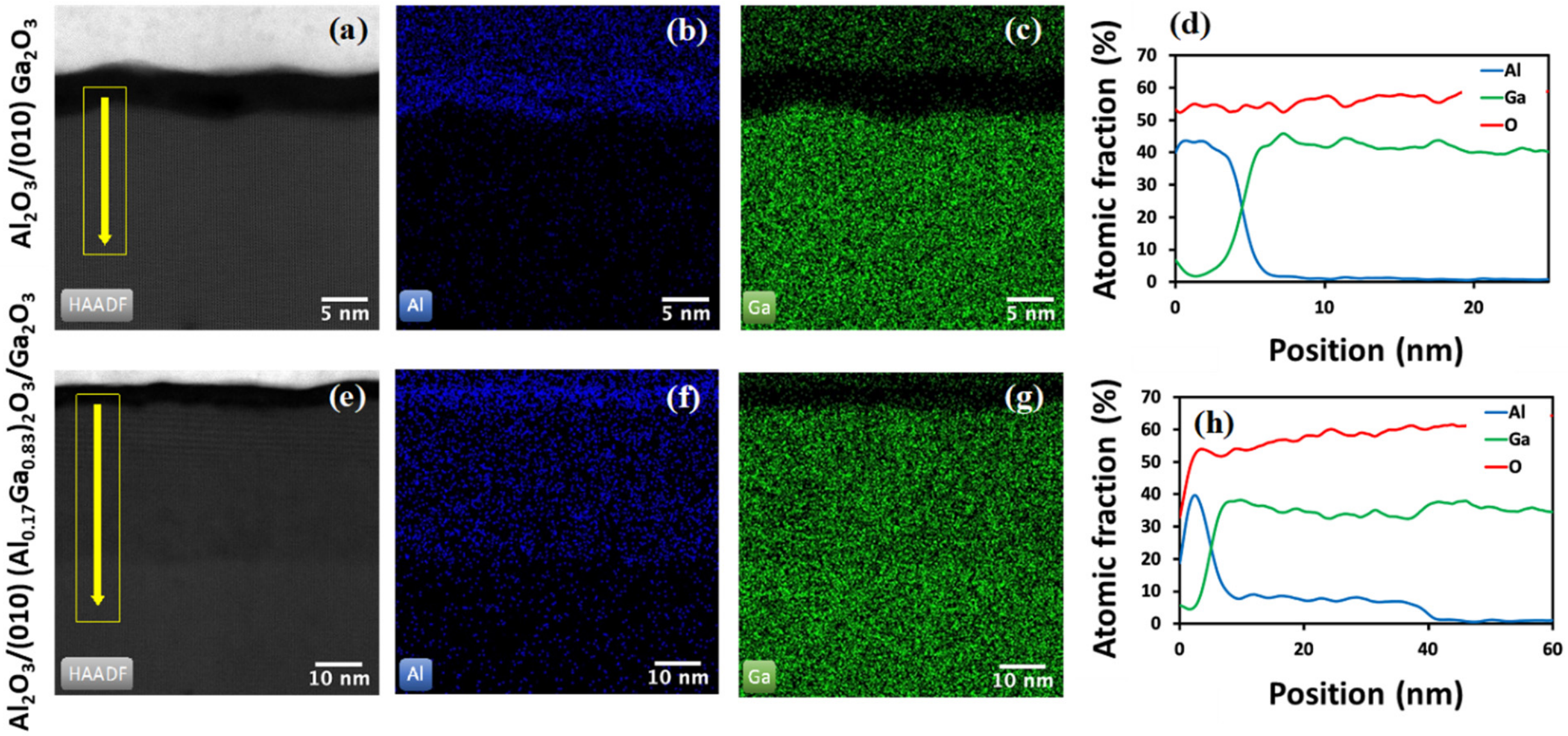

**Figure 5.**

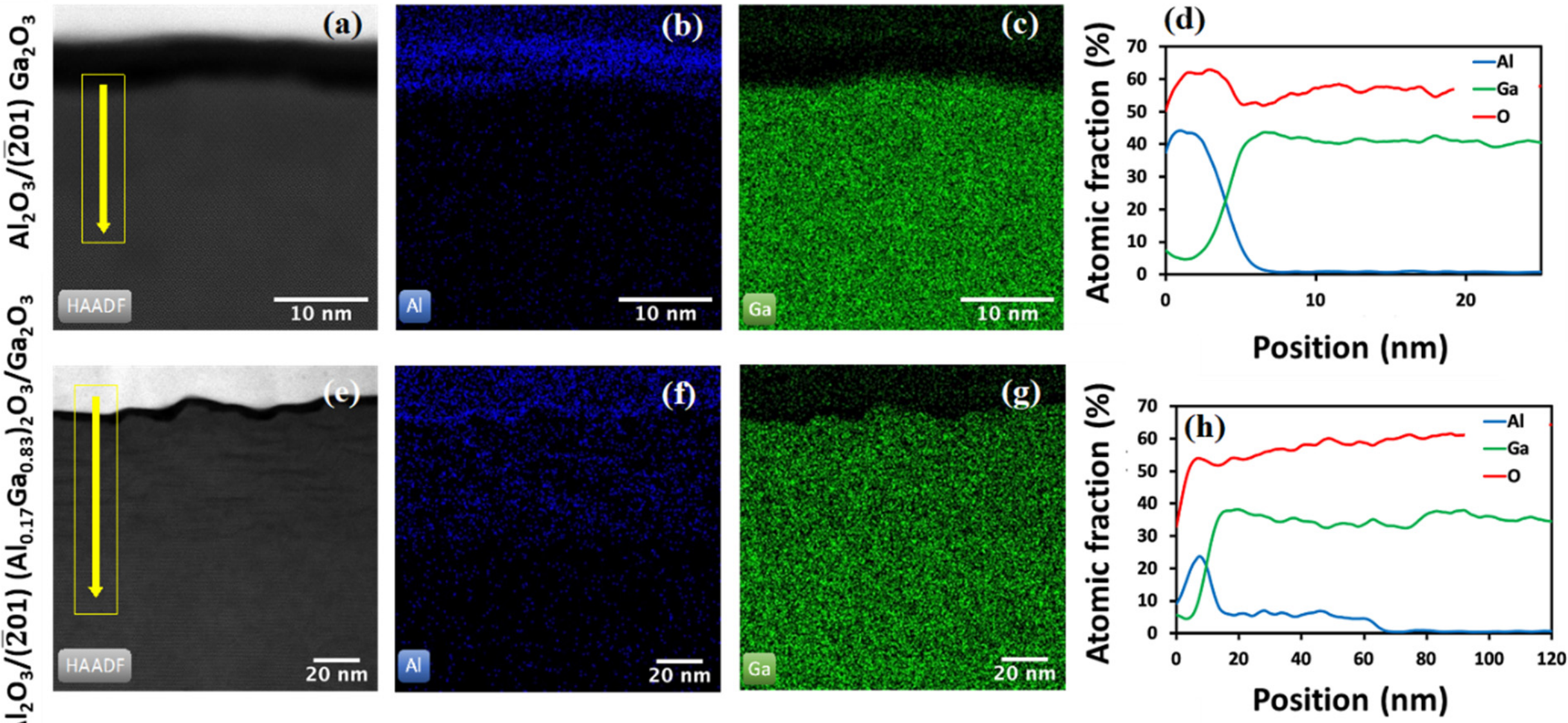

**Figure 6.**

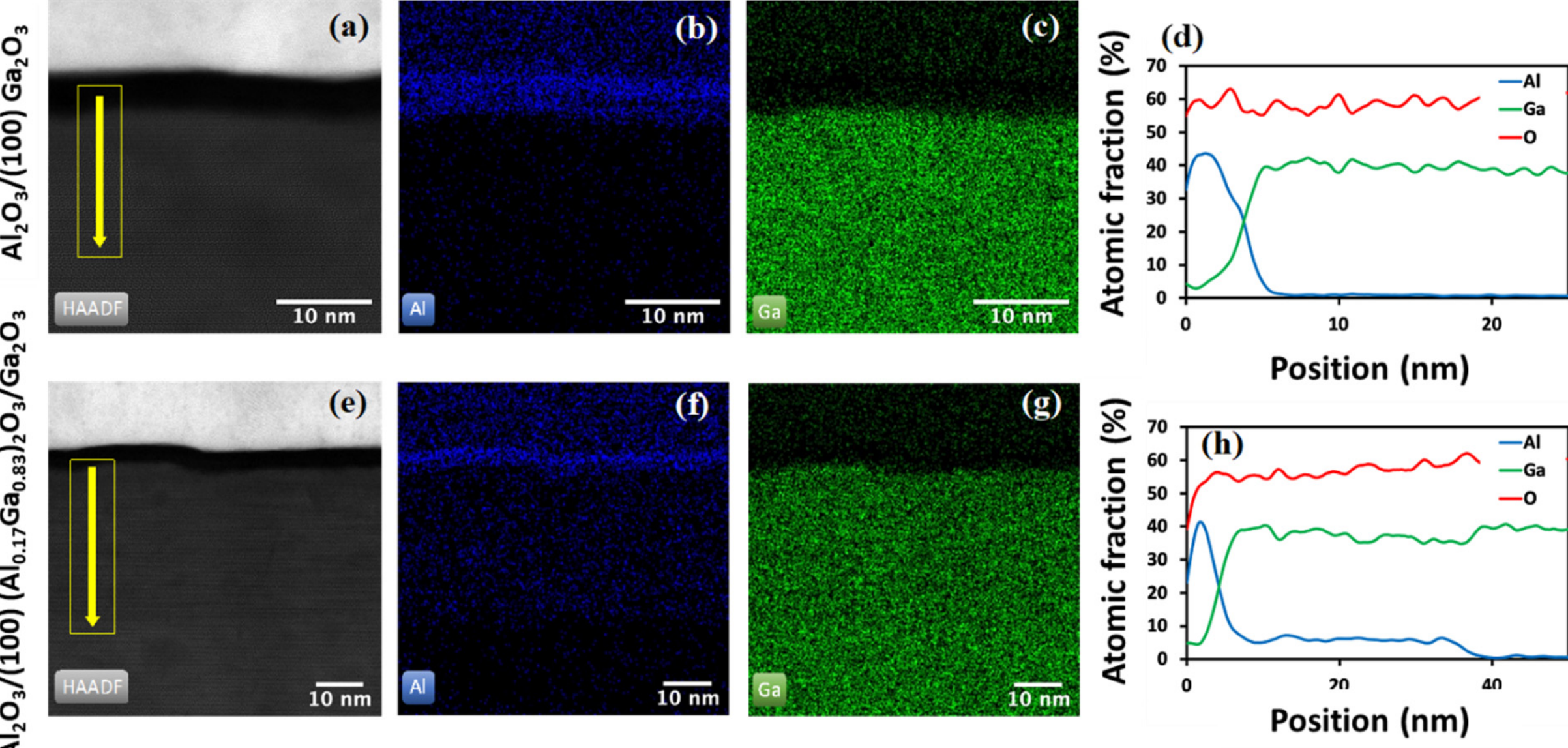

**Figure 7.**

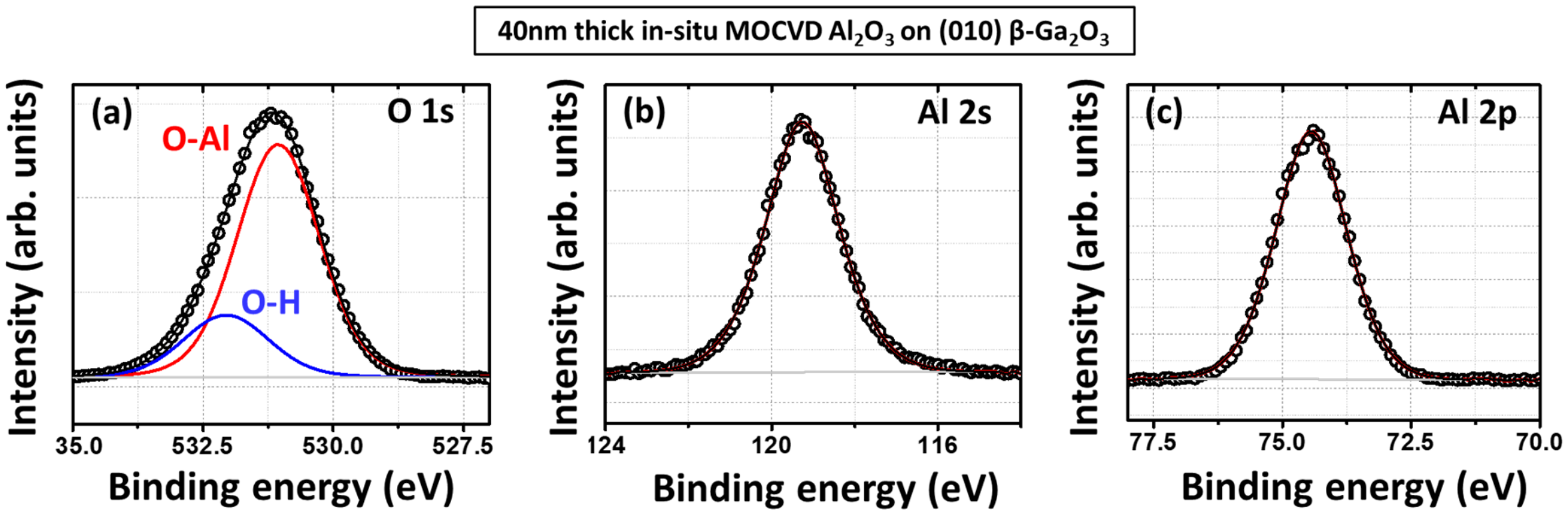

**Figure 8.**

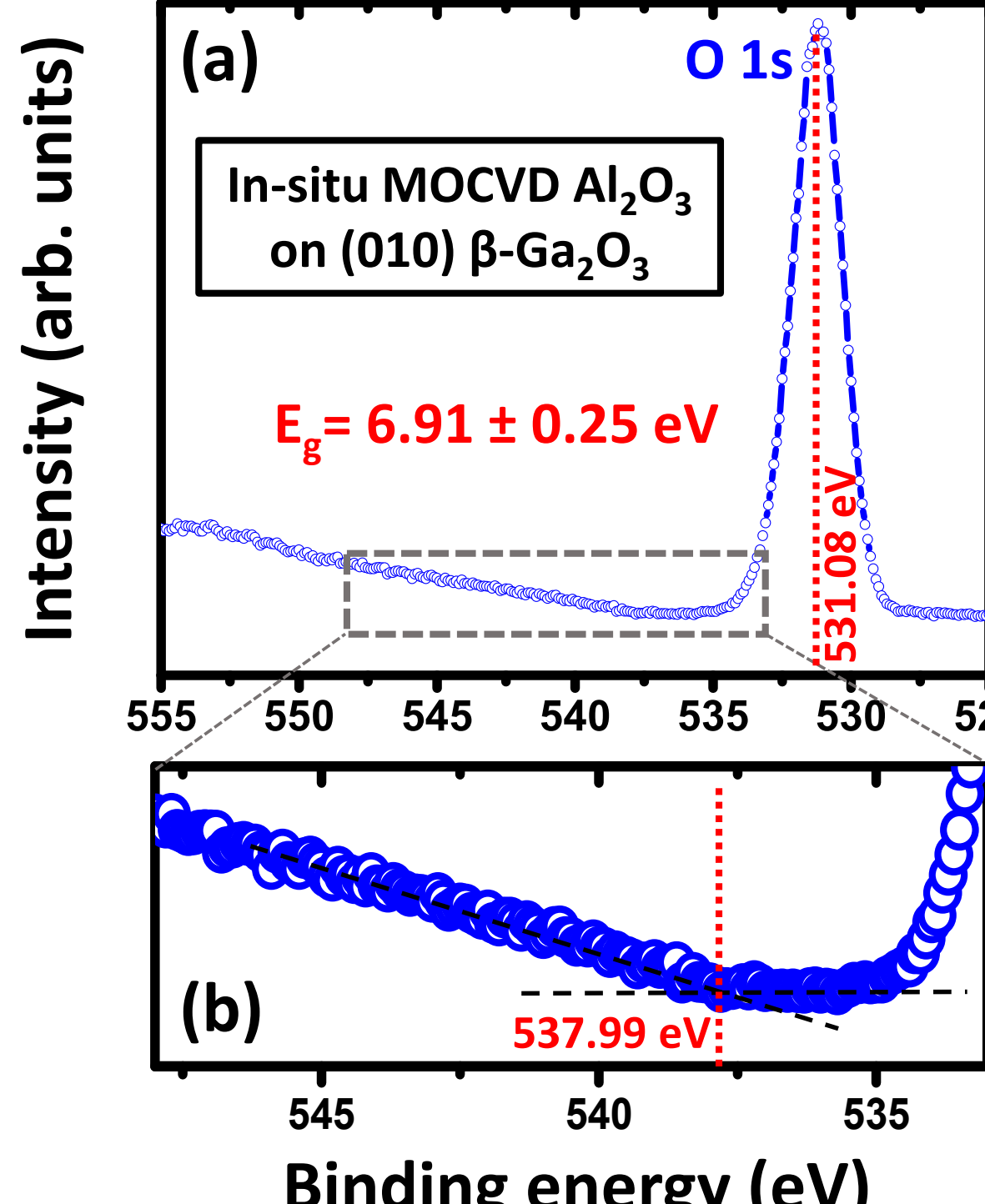

**Figure 9.**

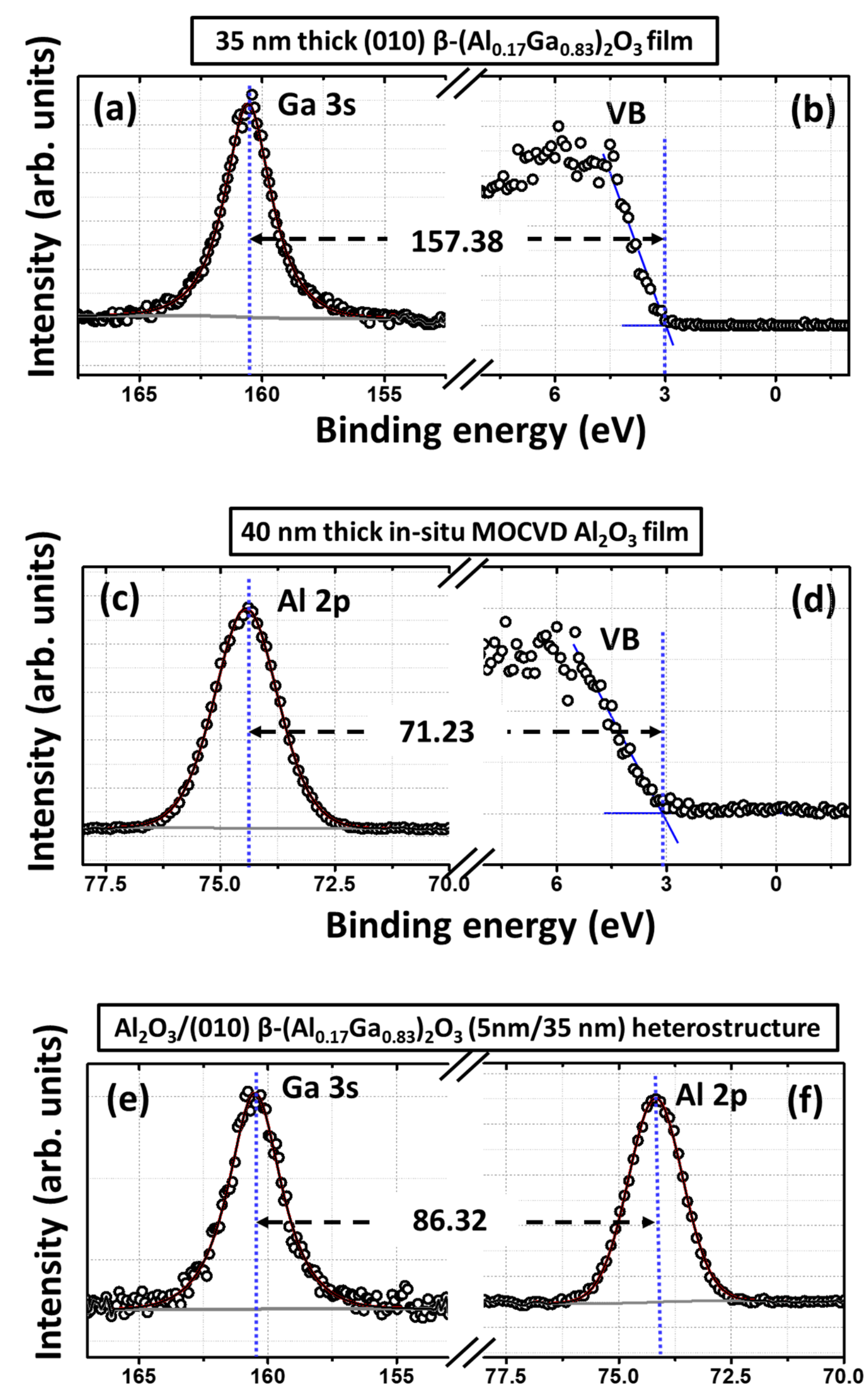

**Figure 10.**

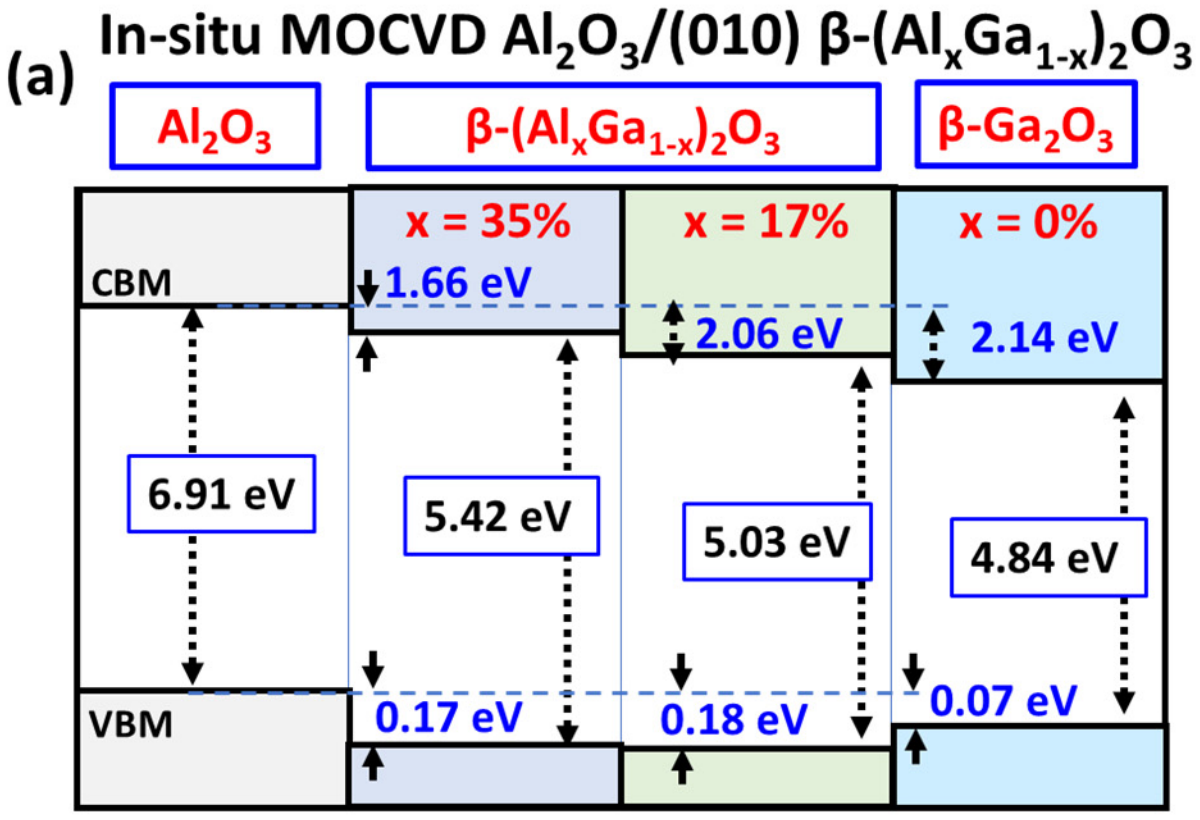

Type II band alignment

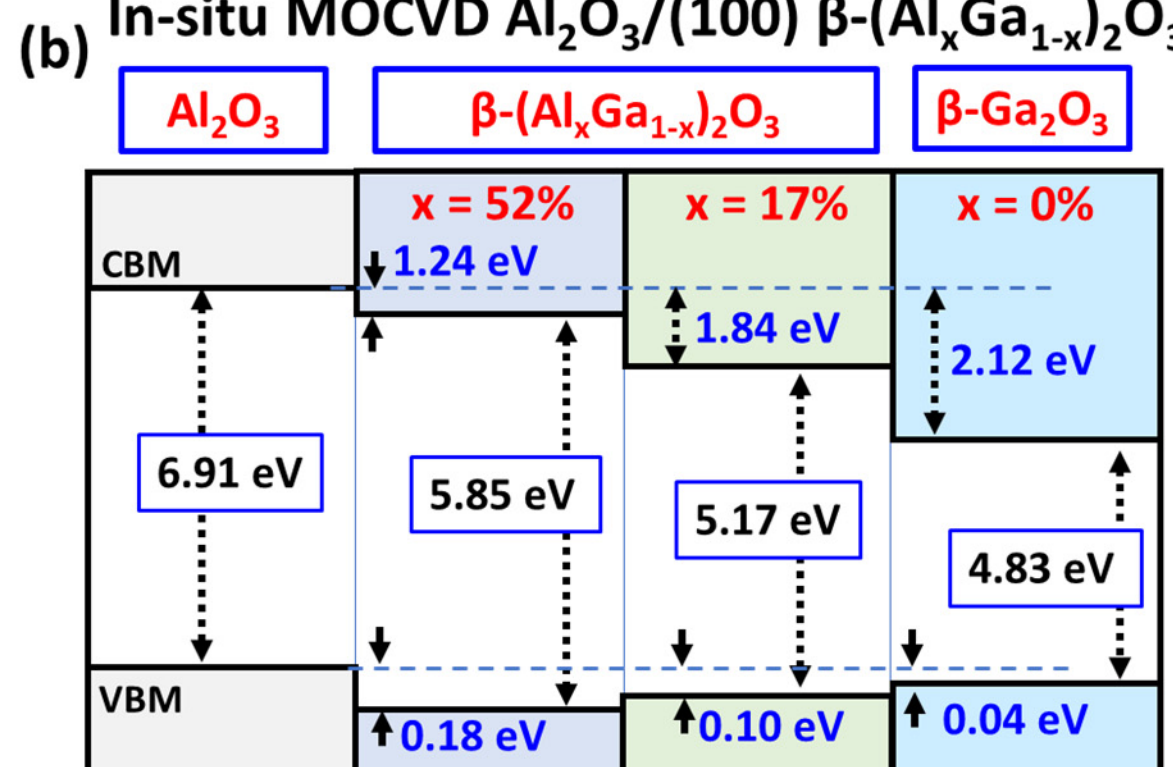

Type II band alignment

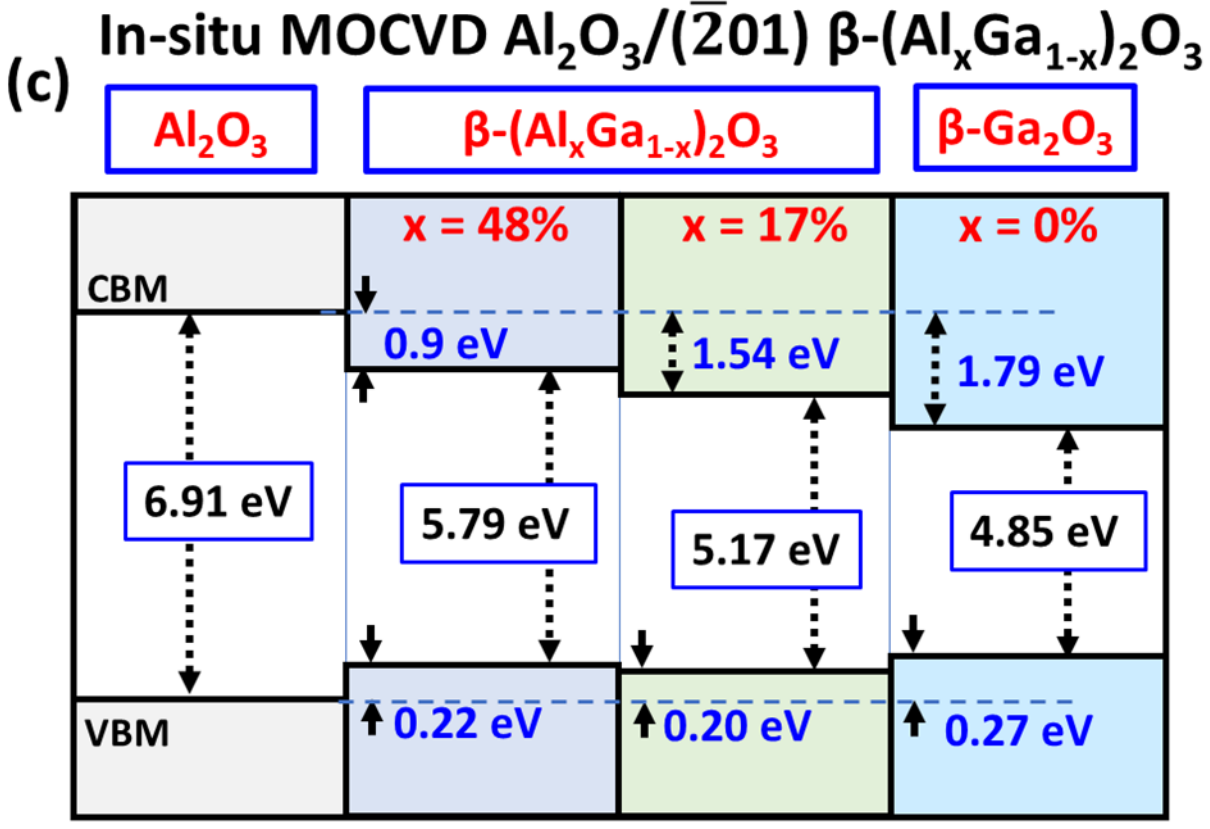

Type I band alignment

**Figure 11.**

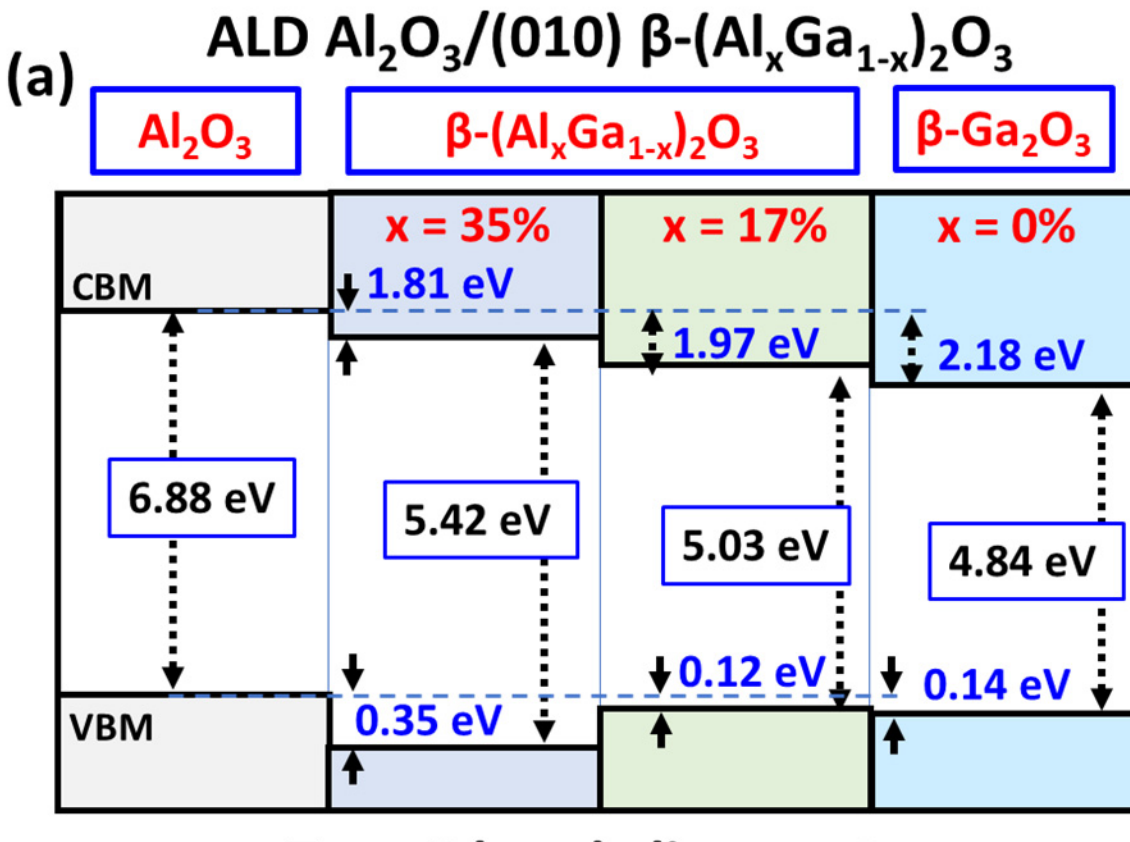

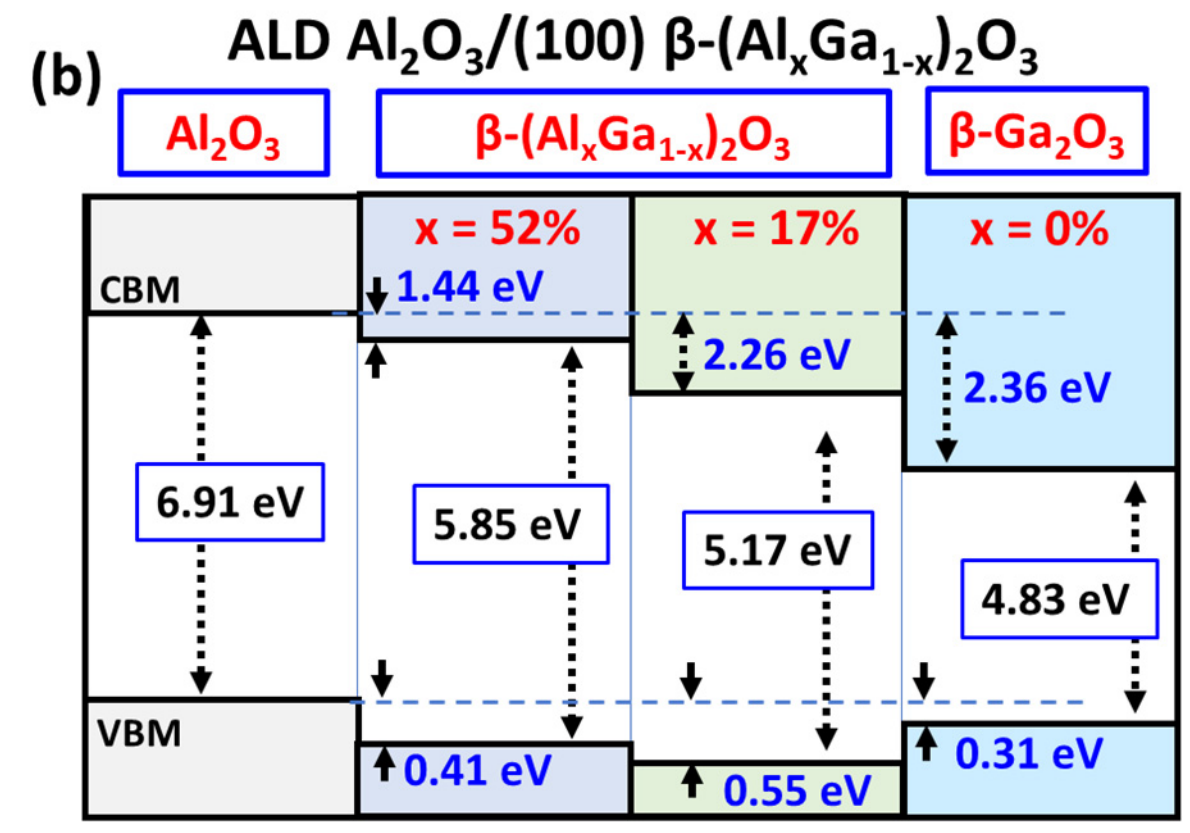

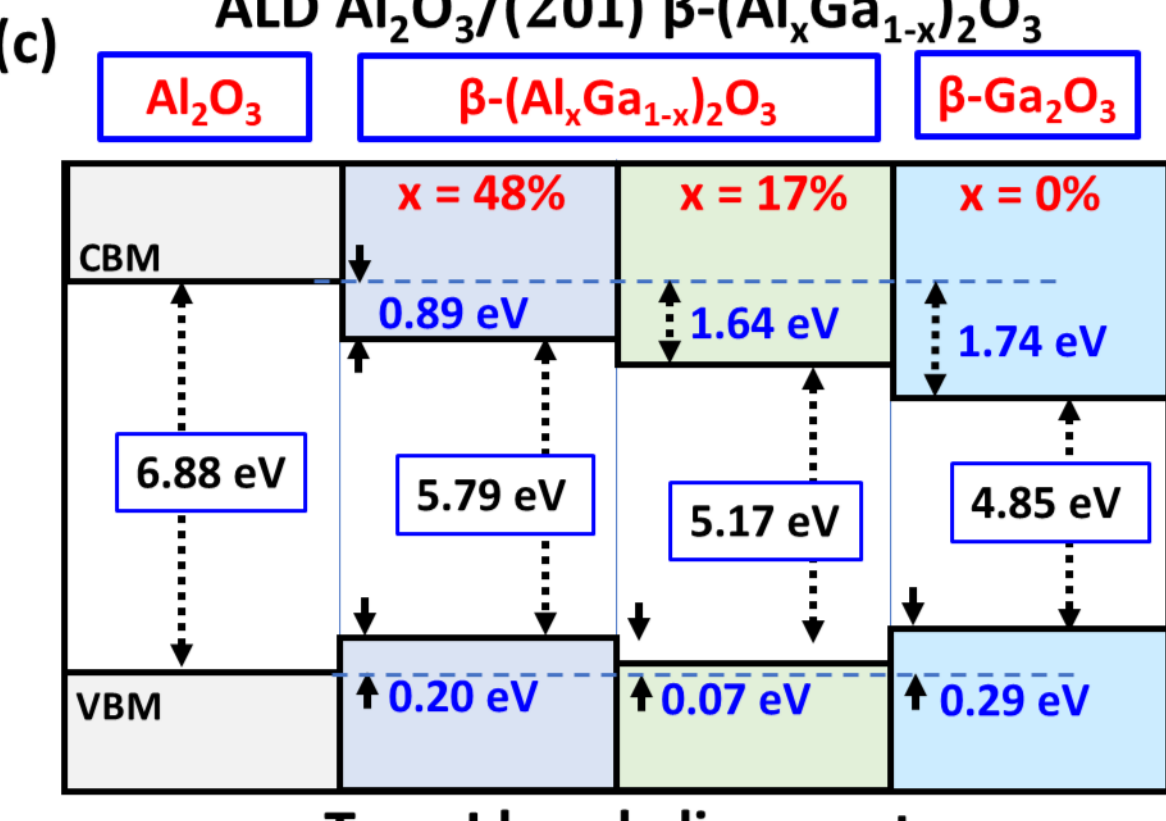

**Figure 12.**

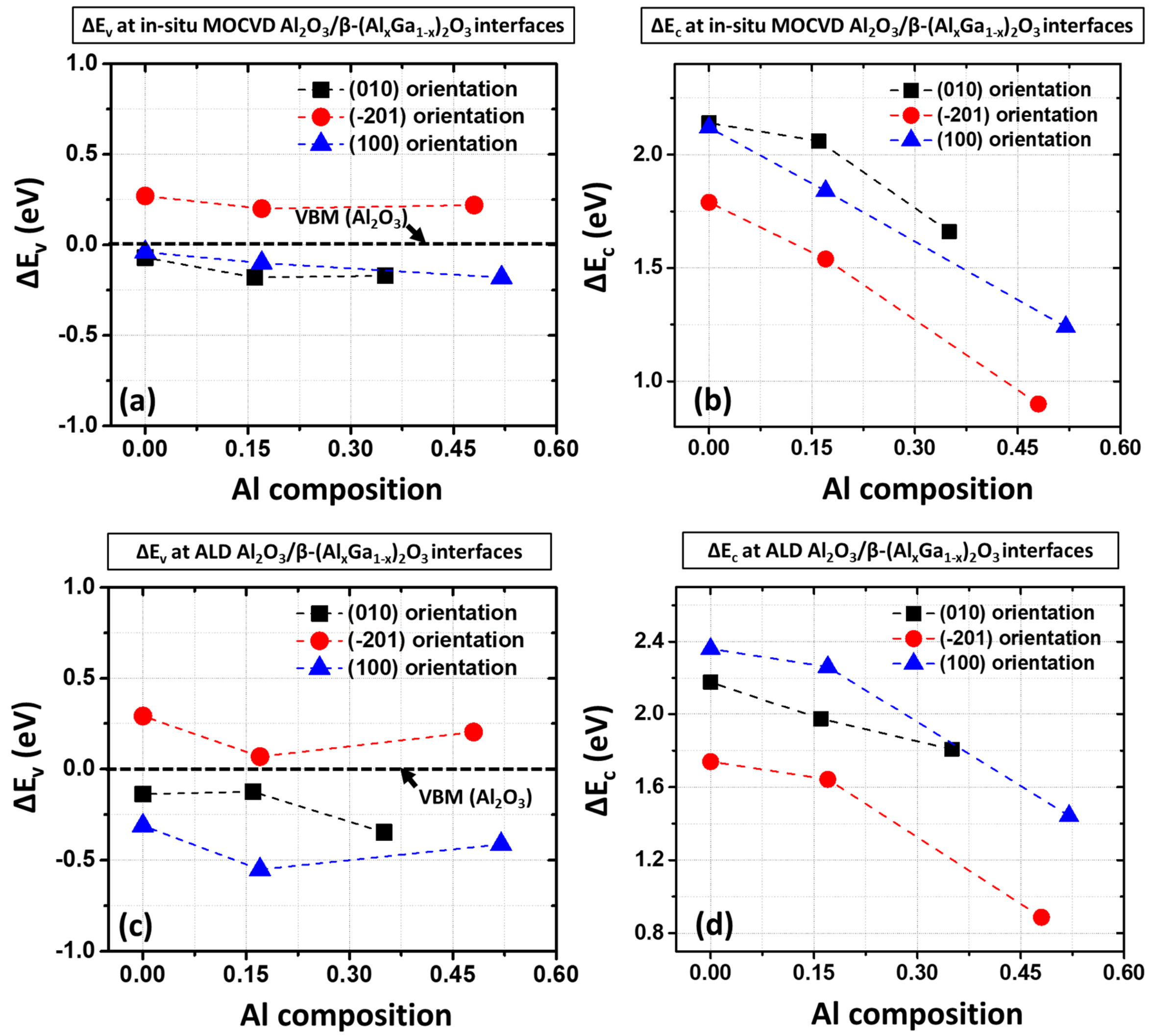